\documentclass[aps, prd, twocolumn, lengthcheck, superscriptaddress, 
nofootinbib]{revtex4-1}

\usepackage{amsmath,amssymb,amsfonts}
\usepackage{slashed}

\usepackage{graphicx}
\usepackage{float} 

\usepackage{verbatim} 
\usepackage{ulem} 
\usepackage{latexsym}
\usepackage{color}
\usepackage{bm}
\def\be{\begin{eqnarray}}
\def\ee{\end{eqnarray}}

\begin{document}

\title{Confined monopoles in a chiral bag}

\author{Fan Lin}
\affiliation{School of Fundamental Physics and Mathematical Sciences,
	Hangzhou Institute for Advanced Study, UCAS, Hangzhou, 310024, China}
\affiliation{Institute of Theoretical Physics, Chinese Academy of Sciences, Beijing 100190,  People’s Republic of China}
\affiliation{University of Chinese Academy of Sciences, Beijing 100049,  People’s Republic of China}
	
\author{Yong-Liang Ma}
\email{ylma@nju.edu.cn}
\affiliation{School of Frontier Sciences, Nanjing University, Suzhou, 215163, China}
\affiliation{International Center for Theoretical Physics Asia-Pacific (ICTP-AP) , UCAS, Beijing, 100190, China}

\date{\today}

\begin{abstract}

The chiral bag model offers a dual description of hadron physics in terms of quarks and hadrons in the sense of Cheshire Cat principle. In this work, we find that, within the chiral bag, confinement is likely caused by monopole condensation. The chiral bag surface can be interpreted as an $\eta'$ domain wall, where a dynamical Chern-Simons theory emerges. Under level-rank duality, the Chern-Simons theory serves as the counterterm introduced to block the so-called color charge leakage. To ensure the correct net baryon number of the full chiral bag, an additional Chern-Simons theory involving the vector meson field arises on the bag surface and extends outside the bag. This leads to a Chern-Simons-Higgs theory localized on the $\eta'$ domain wall, as previously conjectured. We also propose that the skyrmion description of baryons could be understood as a system of monopoles enveloped by a meson cloud.

\end{abstract}

\maketitle


\section{Introduction}

In the low-energy regime, quantum chromodynamics (QCD) predicts color confinement, that is, quarks and gluons are combined to form mesons and baryons. The dynamics of the lightest pseudoscalar mesons, which arise from the chiral symmetry breaking $SU(N_f)_L \times SU(N_f)_R \rightarrow SU(N_f)_V$, dominate strong interaction processes at low energies. This leads to a nonlinear sigma model as the leading order low-energy effective theory of QCD describing pions~\cite{Coleman:1969sm,Callan:1969sn}. Beyond the leading order, additional pion terms can be incorporated into the Lagrangian using a low-momentum expansion, resulting in chiral perturbation theory~\cite{Weinberg:1978kz,Gasser:1983yg}. Baryons can also be included by adding interaction terms consistent with chiral symmetry~\cite{Weinberg:1990rz,Gasser:1987rb}. Although chiral perturbation theory has achieved great success in describing strong interaction processes at low energies, it is insufficient for probing the internal structure of baryons, which has led to  the proposal of various baryon models.

Before the establishment of QCD, Skyrme discovered topological soliton solutions consisting of nonlinearly interacting pion fields, known as skyrmions, and conjectured them as baryons~\cite{Skyrme:1961vq,Skyrme:1962vh,Zahed:1986qz,Atiyah:1989dq,Manton:2004tk,Lee:2003aq}. The rationality of this conjecture was later argued in the large $N_c$ limit, where QCD simplifies and is dominated by planar diagrams~\cite{tHooft:1973alw}. In this limit, baryons can be viewed as solitons composed of interacting mesons since their properties exhibit similar $N_c$ scalings~\cite{Witten:1979kh,Adkins:1983ya,Witten:1983tw}. However, the skyrmion description emphasizes the global properties of baryons rather than their quark composition. The MIT bag model, proposed in the 1970s~\cite{Chodos:1974je,Chodos:1974pn}, fills this gap. Inside the MIT bag, the colored objects, like quarks and gluons, are free or weakly interacting objects, confined by a boundary condition representing the strong attractive force. In the literature, the MIT bag model has been applied to compute baryon masses, magnetic moments, decay rates, and hadron-quark transitions~\cite{Chodos:1974pn,Collins:1974ky,DeGrand:1975cf,Guichon:1987jp,Buballa:2003qv}.

The chiral bag model, an extension of the MIT bag model~\cite{Brown:1979ui,Brown:1979ij,Vento:1980mu,Thomas:1982kv,Rho:1983bh,Chodos:1975ix,Mulders:1984df,Goldman:1980ww}, overcomes the shortcomings of the MIT bag model by incorporating spontaneous chiral symmetry breaking. In this model, the MIT bag is surrounded by a cloud of chiral mesons, such as pions. Quark-meson field interactions occur on the bag surface~\cite{Nadkarni:1985dm}, where the MIT bag description merges with the Skyrme model of baryons. The surface physics is crucial, acting as a bridge between the two descriptions. Moreover, the physics of the chiral bag is expected to be independent of the bag's size or position, a principle known as the Cheshire Cat principle~\cite{Nadkarni:1984eg,Perry:1986sz,Damgaard:1992cy,Nadkarni:1985dn}, which is supported by the numerical check of the flavor-singlet charge of proton~\cite{Lee:1998rj} and reveals the possible hadron–quark continuity at high density~\cite{Ma:2019ery}.

So far, most research has focused on the dynamics of chiral mesons, particularly pions, within the chiral bag model. Less attention has been given to the pseudoscalar isosinglet meson $\eta'$, which gains mass from the $U(1)_A$ anomaly~\cite{Nielsen:1991bk,Nielsen:1992va}. In the large $N_c$ limit, $U(1)_A$ symmetry is restored, and the $\eta'$ becomes a massless Nambu-Goldstone boson. Consequently, the $\eta'$ should not be neglected, especially given recent developments in the physics of $\eta'$ domain walls and the topology of low energy QCD at high density~\cite{Ma:2020nih}. This prompts a re-examination of the $\eta'$'s role in the chiral bag model.

The $U(1)_A$ problem and the associated $\eta'$ physics have been thoroughly investigated over the last century~\cite{Witten:1979vv,Veneziano:1979ec,DiVecchia:1980yfw}. Recently, ’t Hooft anomaly matching has been extended to discrete symmetries and higher-form symmetries~\cite{Gaiotto:2014kfa,Kapustin:2014lwa,Kapustin:2014gua}, offering new insights into the $\eta'$ domain wall. It was found that the $\eta^\prime$ domain wall supports a topological $SU(N_c)_{-N_f}$ Chern-Simons theory~\cite{Gaiotto:2014kfa,Gaiotto:2017yup,Gaiotto:2017tne}, conjectured to be equivalent to the $U(N_f)_{N_c}$ Chern-Simons theory under level-rank duality~\cite{Hsin:2016blu}. In the single-flavor case, baryons can be constructed on the $\eta^\prime$ domain wall similarly to quantum Hall droplets. These Hall droplets have spin $N_c/2$ and chiral modes, carrying the correct baryon number~\cite{Komargodski:2018odf}.

Baryons as quantum Hall droplets can also be interpreted as chiral bags in a (2+1)-dimensional strip, using the Cheshire Cat principle~\cite{Ma:2019xtx}. For small bag radii, the bag reduces to a vortex line—the "smile" of the cat—where gapless quarks flow, all spinning in the same direction. The disk enclosed by the smile is described by an emergent topological field theory due to the Callan-Harvey anomaly inflow~\cite{Callan:1984sa}. For a more concrete description, it has been shown that a Chern-Simons-Higgs theory exists on the $\eta^\prime$ domain wall~\cite{Lin:2023qya}. This theory has vortex solutions, with vortices carrying unit topological charge and naturally spinning $N_c/2$.  These vortices exhibit $N_c$-scaling consistent with baryon properties, suggesting an identification of the two. Through particle-vortex duality, the Zhang-Hansson-Kivelson theory suggests that quarks carry topological charge $1/N_c$ and obey fractional statistics. Furthermore, two-dimensional baryon models have also been realized in nuclear matter through various approaches~\cite{Bigazzi:2022luo,Eto:2023tuu,Huang:2017pqe,Gudnason:2014nba,Eto:2023wul,Fukushima:2018ohd}.

It is worth emphasizing that studies on the chiral bag model with an $\eta'$ boundary condition have revealed color leakage from the bag. This breaks confinement, and a counterterm on the bag surface has been proposed to restore gauge invariance~\cite{Rho:1983bh}. This counterterm corresponds to an $SU(N_c)_{N_f}$ Chern-Simons theory,  but an $SU(N_c)_{-N_f}$ Chern-Simons theory arises on the $\eta'$ domain wall~\cite{Komargodski:2018odf,Ma:2019xtx,Lin:2023qya}. This motivates us to reconsider what happens inside the bag and how it affects surface physics when the chiral bag model includes an $\eta'$ boundary condition, which is the focus of this article.

In this work, we concentrate the Chern-Simons theories existing on the chiral bag surface. Within the chiral bag, confinement arises from monopole condensation, and the bag surface can be interpreted as an $\eta'$ domain wall, where an $SU(N_c)_{N_f}$ Chern-Simons theory emerges to remain 1-form gauge invariance. The same Chern-Simons theory, previously introduced as a counterterm, also prevents color charge leakage~\cite{Rho:1983bh}, and should be identified. To guarantee the correct net baryon number for the chiral bag, a $U(N_f)_{N_c}$ Chern-Simons theory for vector meson fields is introduced on the bag surface and extended to the exterior, which has been used for constructing baryons on the $\eta'$ domain wall~\cite{Komargodski:2018odf,Ma:2019xtx,Lin:2023qya}.

The paper is organized as follows: In Sec.~\ref{sec:II}, we review the color anomaly in the chiral bag model. Secs.~\ref{sec:monople} and~\ref{sec:dyn} discuss the hypothesis that confinement is caused by monopoles confined inside the bag, and shows the Chern-Simons theories exist on the bag surface. The importance of vector meson fields on the bag surface is emphasized, as they ensure the correct baryon number for the chiral bag and give the effective theory whose vortices as baryons on the $\eta'$ domain wall~\cite{Komargodski:2018odf,Ma:2019xtx,Lin:2023qya}. Finally, we conclude our understanding of baryon construction and discuss why soliton solutions of chiral mesons can be regarded as baryons in Sec.~\ref{sec:concl}.

\section{Color anomaly in a chiral bag model} 
\label{sec:II}

As stated in the introduction, the chiral bag model  provides a powerful framework for investigating baryon properties. The chiral bag is defined as a finite spatial region enclosed by a closed surface, within which massless quarks and gluons are confined. Outside this region, a meson cloud composed of various meson fields exists. To consistently connect the interior and exterior of the bag, chiral boundary conditions are imposed on the bag surface~\cite{Nadkarni:1985dm}. If the boundary condition involves only the pseudoscalar isospin triplet meson fields (the pion fields), the vacuum develops a nonzero baryon number~\cite{Rho:1983bh,Goldstone:1981kk}. However, the chiral boundary condition does not confine the baryon current at the quantum level. Time-dependent pion fields allow the baryon number to leak out in the form of a topological current. This topological current aligns with the Skyrme picture, in which the baryon is viewed as a soliton of the pion fields.

When the coupling to the pseudoscalar isosinglet field $\eta^\prime$ is included at the bag  boundary, color charges leak. To enforce total confinement of color charge, an additional boundary term was suggested~\cite{Nielsen:1991bk}. For completeness, we now review the color anomaly in the chiral bag model relevant to this work.

To confine color charge within the bag, the gluons are subject to the following boundary conditions on the bag surface~\cite{Chodos:1974je}
\begin{equation}
	\boldsymbol{n}\cdot\boldsymbol{E}^a_G=0,\quad\boldsymbol{n}\times\boldsymbol{B}^a_G=0, \label{BD1}
\end{equation}
where $\bm{n}$ is the outward normal to the bag surface $\Sigma$, $\boldsymbol{E}^a_G = G^{a,i0}$ is the color electric field tangent to the surface, and $\boldsymbol{B}^a_G = -\frac{1}{2} \epsilon^{ijk} G^a_{jk}$ is the color magnetic field orthogonal to the surface. The indices $a, b, c$ denote color. Under these conditions, quarks can only move along the surface and cannot escape from the bag at the classical level. However, at the quantum level, the situation changes.

One can couple the pseudoscalar isosinglet meson field $\eta^\prime$ to quarks at the bag boundary as follows
\begin{equation}
	\left[\mathrm{i}\gamma\cdot\bm{n}+\mathrm{e}^{\mathrm{i}\gamma_5 {\eta^\prime_\Sigma}/{f_{\eta^\prime}}}\right]\psi_\Sigma=0, 
	\label{Boun}
\end{equation}
where $f_{\eta^\prime}$ is the decay constant of $\eta^\prime$ and the subscript $\Sigma$ specifies that field defined at the bag boundary. For convenience, we later redefine the $\eta^\prime$ field by $\eta^\prime \rightarrow \eta^\prime f_{\eta^\prime}$ to absorb the decay constant, so that $\eta^\prime$ becomes a dimensionless field. Inside the bag region $\Omega$ with boundary $\partial \Omega = \Sigma$, the physical states $ \mid\mathrm{phys} \rangle$ inside the bag must satisfy Gauss' law
\begin{equation}
	(\nabla \cdot \boldsymbol{E}^a - j_0^a) \Theta_\Omega \mid \mathrm{phys} \rangle = 0,
\end{equation}
where $\Theta_\Omega$ is the step function with support on the bag cavity $\Omega$, and $j_0^a$ is the color charge density. The color charge $Q^a_G$ is given by
\be
Q^a_G & = & \int_{\Omega} \mathrm{d}^3x \, j_0^a = \int_{\Omega} \mathrm{d}^3x \left( g\psi^\dagger \frac{1}{2} \lambda^a \psi + g f^{abc} G_i^b E_G^{ci} \right) \nonumber\\
& \simeq & \oint_{\Sigma} \mathrm{d}S \, \boldsymbol{E}^a_G \cdot \bm{n},
\label{eq:ColorCharge}
\ee
where $G_i^b$ is the spatial component of gluon field with color ``$b$" and the last relation follows from Gauss' law on physical states. $\mathrm{d}S$ is the area element of the bag surface $\Sigma$. $\lambda^a$ is the Gell-Mann matrix, and $f^{abc}$ are the structure constants of the $SU(3)$ color algebra. Although the QCD action is invariant under local gauge transformations inside the bag, color charge $Q^a$ is not conserved at the quantum level. Quantitative calculations, based on the multiple reflection expansion of the fermion propagator in cavity quantum chromodynamics~\cite{Hansson:1982cu}, show that interactions between quarks and gluons inside the bag cannot be ignored~\cite{Nielsen:1991bk}. For one-flavor species in the quasi-Abelian case, color charge leaks out when there is time variation in the $\eta^\prime$ field
\begin{equation}
	\frac{\mathrm{d}Q^a_G}{\mathrm{d}\eta^\prime_\Sigma} = \frac{g^2}{8\pi^2} \oint_{\Sigma} \mathrm{d}S \, \boldsymbol{B}^a_G \cdot \bm{n}.
\end{equation}
 Combining this equation with Eq.~\eqref{eq:ColorCharge}, we obtain
\begin{equation}
	\frac{\mathrm{d}Q^a_G}{\mathrm{d}\eta^\prime_\Sigma} = \frac{g^2}{8\pi^2} \oint_{\Sigma} \mathrm{d}S \, \boldsymbol{B}^a_G \cdot \bm{n} \simeq \frac{\mathrm{d}}{\mathrm{d}\eta^\prime_\Sigma} \oint_{\Sigma} \mathrm{d}S \, \boldsymbol{E}^a_G \cdot \bm{n}.
	\label{dQ}
\end{equation}
This equation is also argued to hold for non-Abelian color magnetic fields. This effect essentially constitutes a color anomaly. Assuming a nonzero color magnetic field $B^a$ (in some fixed gauge) in the vicinity of the bag wall pointing perpendicularly to the wall, massless quarks can occupy Landau levels aligned perpendicular to the wall. If the $\eta^\prime$ field is time dependent, quarks in the lowest Landau level are not merely reflected at the wall—they also gain energy and momentum. This ``phase kick" pushes the reflected quarks into or out of the Dirac sea. Since these quarks carry color, this leads to a gain or loss of color charge, resulting in the color anomaly.

The leakage of color charge is problematic as it violates confinement and breaks gauge invariance. To prevent this, a gauge-dependent counterterm has been proposed~\cite{Nielsen:1991bk}
\be
S_{\mathrm{CT}} & = & \frac{g^2 N_f}{16\pi^2} \oint_{\Sigma} \mathrm{d}S \, \eta^\prime_\Sigma \, n_{\mu} K_5^\mu, \nonumber\\
K_5^\mu & = & \epsilon^{\mu\nu\alpha\beta} \left( G_\nu^a G_{\alpha\beta}^a - \frac{2}{3} f^{abc} g G_\nu^a G_\alpha^b G_\beta^c \right).
\label{CT}
\ee

However, there is an implicit assumption in going from Eq.~\eqref{dQ} to Eq.~\eqref{CT}, namely that $\eta^\prime$ is chosen to be zero inside the bag. This allows for the integration of Eq.~\eqref{dQ} from the interior of the bag to the surface, yielding the correct counterterm~\eqref{CT}.  At next-to-leading order in the large $N_c$ expansion, the effective Lagrangian for $\eta^\prime$ is given by~\cite{Witten:1979vv,Veneziano:1979ec}
\begin{equation}
	\mathcal{L}^{\mathrm{eff}}_{\eta'} = \frac{N_f f_\pi^2}{8} \, \mathrm{d}\eta' \wedge \star \mathrm{d}\eta' + \frac{f_\pi^2}{8 N_f} m_{\eta'}^2 \min_{n \in \mathbb{Z}} \left( N_f \eta' + \theta - 2\pi n \right)^2,
	\label{Leta}
\end{equation}
where we observe that $N_f \eta'$ transforms together, giving rise to infinite vacuum branches labeled by $n$. Aside from $\eta^\prime = 0$, other physical vacua exist. In general, the vacuum value $\eta^\prime_{\mathrm{in}}$ inside the bag can be chosen arbitrarily without affecting the local dynamics of quarks and gluons. Regardless of the choice of vacuum, a thin layer near the bag surface must exist where $\eta^\prime_{\mathrm{in}}$ changes sharply to $\eta^\prime_{\Sigma}$.

Upon integrating Eq.~\eqref{dQ} and recomputing the counterterm, a new term emerges
\begin{equation}
	\frac{\eta^\prime_{\Sigma} - \eta^\prime_{\mathrm{in}}}{2\pi} \times \int_{\Sigma} \frac{N_f}{4\pi} \left( G \mathrm{d}G - \mathrm{i} \frac{2}{3} G^3 \right),
	\label{CTCS}
\end{equation}
where differential form notation is used, and the coupling is absorbed into the gauge field. If we simply choose $\eta^\prime_{\Sigma} - \eta^\prime_{\mathrm{in}}=2\pi$, an $SU(N_c)_{N_f}$ Chern-Simons theory clearly arises on the bag surface. This result bears a striking resemblance to recent studies on $\eta^\prime$ domain walls, where an $SU(N_c)_{-N_f}$ Chern-Simon field theory is predicted~\cite{Gaiotto:2017yup,Gaiotto:2017tne}, and which has been used to construct one-flavor baryons in 2+1 dimensions~\cite{Komargodski:2018odf,Ma:2019xtx,Lin:2023qya}.

In fact, if we consider that $\eta^\prime$ may differ from its vacuum inside the bag, the bag surface effectively becomes a domain wall. This motivates an exploration of the relationship between the $SU(N_c)_{N_f}$ Chern-Simons theory on the chiral bag surface and the $SU(N_c)_{-N_f}$ Chern-Simons theory on the $\eta'$ domain wall. In the following sections, we further investigate the chiral bag model by focusing on $\eta'$ physics within the framework of topological field theory.

\section{Confined monopoles inside the bag}
\label{sec:monople}

As reviewed above, the core idea of the chiral bag model is to impose boundary conditions that not only confine quarks and gluons within the bag but also couple consistently to mesonic dynamics outside. One can solve the equations of motion with these boundary conditions to calculate various physical quantities. However, for the baryon number, the situation is more subtle due to its topological nature. Although quarks are confined within the bag, the baryon number of the bag is entirely determined by the boundary conditions on the mesons, or, in other words, by the physics on the bag surface~\cite{Niemi:1984vz}. For the chiral bag with $N_f \ge 2$, the baryon number is related to the Gaussian curvature of pion fields in flavor space~\cite{Goldstone:1983tu}, which is consistent with the skyrmion picture. In such a case, the role of the $\eta^\prime$ meson is often neglected as it is considered irrelevant or integrated out since it is much heavier than pions. However, if we consider $N_f = 1$, how can we obtain a non-trivial baryon number consistent with the baryon spectrum? In this case, only the pseudoscalar isosinglet meson $\eta^\prime$ exists, and the confined quarks inside the bag are essential for a complete description.

To describe a color singlet constituted by quarks inside the bag, a mechanism that causes confinement is necessary. It is worth noting that in Eq.~\eqref{dQ}, color charge is induced when a nonzero color magnetic field crosses the surface with a constant $\eta^\prime$. Nonzero color magnetic fields may stem from vacuum fluctuations, but these are tiny in a stable system. We can also conjecture the existence of monopoles inside the bag, which would generate a significant color magnetic field on the bag surface. Since gluons belong to the $SU(N_c)$ Yang-Mills theory, we can assume quarks behave like monopoles inside the bag, and the condensation of $N_c$ monopoles leads to confinement. It is suggested that the classification of monopole charges follows the discrete group $Z_{N_c}$, the center of the gauge group $SU(N_c)$. The topological properties of $SU(N_c)$ gauge theory can then be described within the framework of topological field theory~\cite{Dierigl:2014xta,Banks:2010zn,Gukov:2013zka}. Using a magnetic Abelian-Higgs model, the discrete gauge group $Z_{N_c}$ can be embedded in a $U(1)$ theory.

To describe the monopoles confined within the bag, we follow the method proposed in~\cite{Dierigl:2014xta}. Consider a complex scalar field $\Phi$ that carries $N_c$ magnetic charges, with $\tilde{A}$ representing the magnetic dual field of the usual gauge field $A$ in $U(1)$ theory. The covariant derivative is written as,
\begin{equation}
	D\Phi=(\mathrm{d}-iN_c\tilde{A})\Phi.
\end{equation}
The complex scalar field $\Phi$ represents the density of monopoles, which should have a Higgs-type potential resulting in a nonzero vacuum expectation value $\left\langle\left|\Phi\right|\right\rangle\equiv v>0$. In the low-energy region, $\Phi=\upsilon e^{i\varphi}$, so the action is dominated by pure gauge configurations satisfying
\begin{equation}
	D\Phi=i(\mathrm{d}\varphi-N_c\tilde{A})v \quad\Rightarrow\quad \mathrm{d}\varphi-N_c\tilde{A}=0,
\end{equation}
which precisely describes the discrete $Z_{N_c}$ gauge theory. By dualizing the magnetic field $\tilde{A}$ to return to the electric description via the usual gauge field $A$, one obtains the action of the so-called BF-theory~\cite{Horowitz:1989ng}
\begin{equation}
	S=\frac{i}{2\pi} \int{ \mathrm{d}\tilde{A} \wedge (\mathrm{d}A - N_c B)}.
\end{equation}
This action possesses a 1-form gauge symmetry parametrized by a 1-form $\lambda$, which fulfills the quantization condition over a closed two-surface---the bag surface $\Sigma$---as
\begin{equation}
	\frac{1}{2\pi}\oint \mathrm{d}\lambda\in\mathbb{Z},\quad A\to A+N_c\lambda, \quad B\to B+\mathrm{d}\lambda.
\end{equation}
This indicates that the added magnetic charge $\Delta m$ into the bag must be a multiple of the condensed monopole charge $N_c$
\begin{equation}
	m=\frac1{2\pi}\oint \mathrm{d}A\to m+\frac1{2\pi}\oint N_c \mathrm{d}\lambda=m+jN_c, \quad j\in \mathbb{Z}.
	\label{Char}
\end{equation}

Now, let us consider what happens when the $\eta^\prime$ field crosses the bag surface. This situation arises from the transformation similarity between the $\eta^\prime$ field and $\theta$, leading to vacuum jumps between different branches. As shown in Eq.~\eqref{Leta}, $\eta^\prime$ and $\theta$ together share multiple vacua labeled by \( n \), resulting in the existence of domain walls. In the \( SU(N_c) \) Yang-Mills theory with a \( \theta \) term, 't Hooft anomaly matching requires that each vacuum label \( n \) is accompanied by a topological term in the action~\cite{Kitano:2020evx}
\begin{equation}
	S = \frac{i}{2\pi} \int d\tilde{A} \wedge (dA - N_c B) + \frac{N_c \theta}{4\pi} B \wedge B - \frac{N_c n}{2} B \wedge B.
\end{equation}	
To derive the interaction between $\eta^\prime$ and the gauge fields \( A \) and \( B \), we perform the shift \( \theta \to \theta + N_f \eta^\prime \) and then set \( \theta = 0 \). The action becomes
\begin{equation}
	S = \frac{i}{2\pi} \int d\tilde{A} \wedge (dA - N_c B) + \frac{N_c N_f \eta^\prime}{4\pi} B \wedge B - \frac{N_c n}{2} B \wedge B.
\end{equation}
If we naively set \( \eta_{\mathrm{in}}^\prime = 0 \) inside the bag and \( \eta_{\Sigma}^\prime = 2\pi \) outside, the bag boundary effectively behaves as an $\eta^\prime$ domain wall, where \( N_f \eta^\prime \) undergoes a \( 2\pi N_f \) shift. Since a shift of \( \Delta \eta^\prime = 2\pi \) corresponds to a topological shift of \( \Delta n = N_f \), we can set \( \eta^\prime = 0 \) and interpret the bag surface as a domain wall with vacuum branches jumping by \( \Delta n = N_f \). Consequently, the bag surface separates two different vacuums, and the effective action is not invariant under 1-form transformation
\be
\Delta S & = & {} -\frac{i}{2\pi}\int \mathrm{d}\left[n \lambda\wedge \mathrm{d}A+\frac{nN_c}{2}\lambda\wedge \mathrm{d}\lambda\right]\nonumber\\
& &{} +\frac{iN_c}{4\pi}\int \mathrm{d}n\wedge(2\lambda\wedge B+\lambda\wedge \mathrm{d}\lambda).
\ee
The first term is a total derivative and does not contribute here since our bag surface is a closed surface. However, the second term develops a contribution on the bag surface
\begin{equation}
	\Delta S_{\mathrm{surface}}=-\frac{iN_c}{4\pi}\int_{\Sigma}(2\lambda\wedge B+\lambda\wedge \mathrm{d}\lambda).
\end{equation}

\begin{figure}
	\centering
	\includegraphics[width=0.9\linewidth]{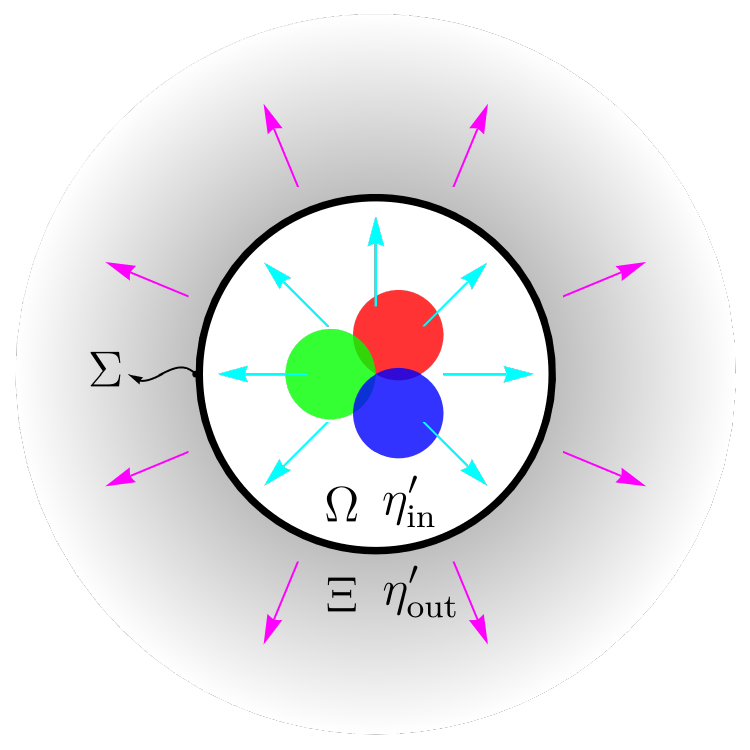}
	\caption[]{Schematic structure of the chiral bag: The chiral bag surface $\Sigma$ acts as an $\eta'$ domain wall, separating space into an interior region $\Omega$ with $\eta_{\mathrm{in}}$ and an exterior region $\Xi$ with $\eta_{\mathrm{out}}$. Inside the bag, color confinement arises from monopole condensation. However, when the color-magnetic flux (shown as cyan arrows) attempts to cross $\Sigma$, it interacts with the $\eta'$ field and would naively lead to color charge leakage. This leakage is precisely cancelled by the effective surface action $S_\Sigma$ [see Eq.~\eqref{eqSSigma}], which encodes a $SU(N_c)_{N_f}$ Chern-Simons theory on the domain wall. This action restores confinement through anomaly inflow. In addition, $S_\Sigma$ also contains interaction $SU(N_c)_{-N_f}$ Chern-Simons theory involving the vector meson field, which propagates into the exterior region. The associated magnetic flux (depicted as magenta arrows) ensures the correct realization of the baryon number outside the bag. }
	\label{fig:bag}
\end{figure}

To maintain gauge invariance, the dynamics on the bag surface must be nontrivial. According to arguments from $\mathcal{N} = 1$ supersymmetric Yang-Mills theory~\cite{Gaiotto:2017yup}, it is suggested that on the bag surface, there exists an $SU(N_c)_{N_f}$ Chern-Simons theory. Due to level-rank duality, an $SU(N_c)_{N_f}$ Chern-Simons theory is identified with a $U(N_f)_{-N_c}$ Chern-Simons theory, leading to the effective action proposed in~\cite{Kitano:2020evx}
\begin{equation}
	S_\Sigma= -\frac{i}{4\pi}\int_{\Sigma}\left[N_c{\rm tr}\left(\mathbb{A}\mathrm{d}\mathbb{A}-i\frac{2}{3}\mathbb{A}^3\right) + 2{\rm tr}(\mathbb{A})\mathrm{d}A\right], \label{eff1}
\end{equation}
where $\mathbb{A}$ is a 1-form $U(N_f)$ gauge field that transforms as
\begin{equation}
	\mathbb{A}\rightarrow \mathbb{A} - \lambda \bm{1}_{N_f \times N_f}.
\end{equation}
It is straightforward to verify that, under the 1-form gauge transformation, the contribution from the bag surface $S_\Sigma$ exactly offsets $\Delta S_{\mathrm{surface}}$. 

It is seen that the gauge field in the bulk $A$ interacts with the dynamical field $\mathbb{A}$ on the bag surface, so $A$ behaves like a background field for the dynamical bag surface. The flux of $\mathbb{A}$ is induced by $A$. Explicitly applying Gauss's law for $\mathbb{A}$ on the bag surface yields
\begin{equation}
	\mathrm{d}\mathbb{A}-i\mathbb{A}^2=-\frac{\mathrm{d}A}{N_c}\bm{1}_{N_f \times N_f}.
\end{equation}
Thus, the vector field $\mathbb{A}$ acquires magnetic flux
\begin{equation}
	\frac{1}{2\pi N_f}\int_\Sigma {\rm tr}(\mathrm{d}\mathbb{A})=-\frac{1}{2\pi N_c}\int_\Sigma \mathrm{d}A,
	\label{Flux}
\end{equation}
indicating that only the $U(1)$ part of $\mathbb{A}$ acquires flux. This flux has the same magnitude as $A$ but with opposite sign, which underlies the dynamical mechanism of charge leakage and blocking.

\section{Dynamics of topological action}
\label{sec:dyn}

We have established that if confinement is assumed to be caused by monopoles, these monopoles can be placed inside the bag. The topological properties can be described by a $Z_{N_c}$ gauge field $A$, whose flux measures the confined monopoles. On the bag surface, a Chern-Simons theory $U(N_f)_{-N_c}$ is induced with respect to the new dynamical field $\mathbb{A}$, which also carries nonzero flux. However, as shown in Eq.~\eqref{Flux}, the corresponding flux has an opposite sign, suggesting that the bag surface must contain antimonopoles. This arrangement effectively seals off the color leak for the chiral bag model, as shown in Eq.~\eqref{CTCS}.

Now, let us focus on the number of monopoles inside the bag and on the bag surface. In the previous section, we set $\eta_{\mathrm{in}}' = 0$ inside the bag and $\eta_{\Sigma}' = 2\pi$ outside. For a more general bag model boundary condition, we will consider both $\eta_{\Sigma}'$ and $\eta_{\mathrm{in}}'$ to be arbitrary. The effective theory on the bag surface, as shown in Eq.~\eqref{eff1}, should be modified as follows
\be
S_\Sigma & = &{} -\left(\frac{\eta_{\Sigma}'-\eta_{\mathrm{in}}'}{2\pi}\right) \nonumber\\
& &{}\times \frac{i}{4\pi}\left[N_c{\rm tr}\left(\mathbb{A}\mathrm{d}\mathbb{A}-i\frac{2}{3}\mathbb{A}^3\right) + 2{\rm tr}(\mathbb{A})\mathrm{d}A\right],
\label{CS}
\ee
where the flux relation from Gauss's law, Eq.~\eqref{Flux}, remains unchanged, but the shift in $\eta'$ introduces the Witten effect~\cite{Witten:1979ey}. The monopoles inside the bag carry the magnetic field $B_A = \mathrm{d}A/N_c$, and when this passes through the bag surface, $B_A$ interacts with $\eta'$, generating the electric charge $Q_A$
\begin{equation}
	\frac{\mathrm{d}Q_A}{\mathrm{d}\eta^\prime_\Sigma}=\frac{\mathrm{d}}{\mathrm{d}\eta^\prime_\Sigma}\oint_{\Sigma}\mathrm{d}S\boldsymbol{E}_A\cdot\bm{n}=\frac{1}{8\pi^2}\oint_\Sigma\mathrm{d}S\boldsymbol{B}_A\cdot\bm{n}, \label{leak1}
\end{equation}
where we have implicitly fixed $\eta_{\mathrm{in}}'$.

Meanwhile, the dynamical field $\mathbb{A}$ on the bag surface also possesses nonzero flux due to Gauss's law, leading to the corresponding electric charge $Q_\mathbb{A}$
\begin{equation}
	\frac{\mathrm{d}Q_\mathbb{A}}{\mathrm{d}\eta^\prime_\Sigma}=\frac{\mathrm{d}}{\mathrm{d}\eta^\prime_\Sigma}\oint_{\Sigma}\mathrm{d}S \boldsymbol{E}_\mathbb{A}\cdot\bm{n}=\frac{1}{8\pi^2}\oint_\Sigma\mathrm{d}S \boldsymbol{B}_\mathbb{A}\cdot\bm{n},
\end{equation}
where $B_\mathbb{A}={\rm tr}(\mathrm{d}\mathbb{A})/N_f$. According to the flux relation in Eq.~\eqref{Flux}, we find
\begin{equation}
	\frac{\mathrm{d}Q_A}{\mathrm{d}\eta^\prime_\Sigma}+	\frac{\mathrm{d}Q_\mathbb{A}}{\mathrm{d}\eta^\prime_\Sigma}=0.
\end{equation}

We observe that $B_\mathbb{A}$ corresponds to the $U(1)$ part in flavor space associated with baryon number conservation, meaning that both the electric charges $Q_A$ and $Q_\mathbb{A}$ correspond to quark number or baryon number. Therefore, we have a nonzero $\mathrm{d}Q_A$ baryon number flowing out from the bag whenever $\eta'_\Sigma$ is time dependent, similar to the color charge leakage discussed in Sec.~\ref{sec:II}. Fortunately, the topological field theory on the bag surface contributes the same quantity but with the opposite sign for the baryon charge, $\mathrm{d}Q_{\mathbb{A}}$, effectively blocking the leak! 

In fact, the dynamical field $\mathbb{A}$ on the bag surface was originally proposed to preserve gauge invariance, which certainly implies charge conservation. On the other hand, multiples of $N_c$ monopoles condensing lead to confinement and must carry color charge. Here, our monopoles behave like quarks, though we do not identify them as such. The nature of baryon number carriers remains unclear; both quarks and gluons are possible candidates, and all baryon numbers we consider are induced. The leak of baryon number is accompanied by a leak of the color charge, and we can simply replace the baryon number charge with color charge in Eq.~\eqref{leak1}
\be
\frac{\mathrm{d}Q_G^a}{\mathrm{d}\eta^\prime_\Sigma} & = & \frac{\mathrm{d}}{\mathrm{d}\eta^\prime_\Sigma}\oint_{\Sigma}\mathrm{d}S \boldsymbol{E}_G^a\cdot\bm{n}=\frac{1}{8\pi^2}\oint_\Sigma\mathrm{d}S \boldsymbol{B}_G^a\cdot\bm{n}, \label{leak1}
\ee
which is precisely the color charge leak found in Eq.~\eqref{dQ} after the coupling constant $g$ is absorbed. Similarly, the Chern-Simons field theory $U(N_f)_{-N_c}$ on the bag surface, dual to $SU(N_c)_{N_f}$, contributes the same quantity but with the opposite sign for the color charge, ${}-\mathrm{d}Q_G^a$, and effectively offsets the color charge leak shown in the above equation. Thus, the Chern-Simons theory in Eq.~\eqref{CS}, initially introduced to maintain gauge invariance, also serves the role of a counterterm as seen in Eq.~\eqref{CTCS}. Under level-rank duality, there are two forms of the topological theory on the surface, providing us with two different descriptions of the same physical phenomenon.

We have seen that, on one hand, the Chern-Simons theory $SU(N_c)_{N_f}$ on the bag surface effectively blocks color charge leakage, thereby restoring gauge invariance and ensuring confinement. On the other hand, the dual Chern-Simons theory $U(N_f)_{-N_c}$ also introduces quark number, with the same quantity but opposite sign relative to the quarks inside the bag. Adding these two contributions together yields a chiral bag model for baryons, albeit with zero baryon number. Thus, an additional mechanism must exist to generate baryon number for the chiral bag.

If we set $\eta_{\mathrm{in}}' = 0$ inside the bag and $\eta_{\Sigma}' = 2\pi N_f$ outside, the theory on the bag surface becomes independent of the choice of interior, as shown in \cite{Kitano:2020evx}. One can consider introducing additional fields on the bag surface to ensure the correct baryon number. We have focused on the interior of the chiral bag, where the quarks can be described by monopoles. The exterior of the chiral bag is also interesting, as it contains many meson fields acting as background fields. 

On the bag surface, a dynamical $U(N_f)_{-N_c}$ Chern-Simons vector field $\mathbb{A}$ exists, indicating that the vector meson fields in flavor space behave similarly. The vector meson field $V$ can couple with the dynamical $\mathbb{A}$ on the bag surface~\cite{Kitano:2020evx}
\be
S_\Sigma & = &{} -\left(\frac{\eta_{\Sigma}' - \eta_{\mathrm{in}}'}{2\pi}\right) \nonumber\\
& &{} \times \frac{i}{4\pi}\int_{\Sigma}\bigg[N_c{\rm tr}\left(\mathbb{A}\mathrm{d}\mathbb{A}-i\frac{2}{3}\mathbb{A}^3\right) \nonumber\\
& &{} \qquad\qquad\; - N_c{\rm tr}\left(V\mathrm{d}V - i\frac{2}{3}V^3\right) \nonumber\\
& &{} \qquad\qquad\; + 2({\rm tr}(V) + {\rm tr}(\mathbb{A}))\mathrm{d}A\bigg].
\label{eqSSigma}
\ee
Therefore, the vector meson fields reside on the bag surface as a $U(N_f)_{N_c}$ Chern-Simons theory and transform as 
\begin{equation}
	V \rightarrow V - \lambda \bm{1}_{N_f \times N_f}.
\end{equation}
Thus, the action is invariant under both 1-form and zero-form gauge transformations. 

Applying Gauss's law to $V$, we obtain a flux relation similar to Eq.~\eqref{Flux}
\begin{equation}
	\frac{1}{2\pi}\int_\Sigma {\rm tr}(\mathrm{d}V) = \frac{N_f}{2\pi N_c}\int_\Sigma \mathrm{d}A.
	\label{FluxV}
\end{equation}
Focusing on the $U(1)_B$ component of the vector meson field, $V_B = {\rm tr}(V)/N_f$, we find that the Witten effect~\cite{Witten:1979ey} on $V$ actually yields the baryon number $Q_\mathrm{in}$
\be
Q_\mathrm{in} & = &{} \left(\frac{\eta_{\Sigma}' - \eta_{\mathrm{in}}'}{2\pi}\right) \times \frac{1}{2\pi}\int_\Sigma {\rm tr}(\mathrm{d}V)\nonumber\\
& = & N_f\left(\frac{\eta_{\Sigma}' - \eta_{\mathrm{in}}'}{2\pi}\right) \times \frac{1}{2\pi N_c}\int_\Sigma \mathrm{d}A.
\ee
For simplicity, consider $N_f = 1, \eta_{\mathrm{in}}' = 0, \eta_{\Sigma}' = 2\pi$ and $\frac{1}{2\pi}\int_\Sigma \mathrm{d}A = N_c$. This represents a chiral bag model for a one-flavor baryon. In the exterior of the bag, the flux of the vector meson field $V$, induced on the surface, continues to extend to infinity or  terminates on another chiral bag as an antibaryon. 

To connect the theory on the bag surface to the outside, $\eta'$ must couple to the vector meson. The effective interaction term concerning the meson field is given by
\begin{equation}
	i\frac{N_c}{8\pi^2}\int \eta' {\rm tr}(\mathrm{d}V \mathrm{d}V).
\end{equation}
This term arises from the hidden Wess-Zumino term, with related physics discussed in~\cite{Karasik:2020zyo,Karasik:2020pwu,Karasik:2022tmd}. The baryon number outside the bag, $N_\mathrm{out}$, is also proposed as
\begin{equation}
	Q_\mathrm{out} = \int_\Xi \left(\frac{\mathrm{d}\eta'}{2\pi}\right) \times \frac{1}{2\pi} \mathrm{d}{\rm tr}(V),
\end{equation}
where the integral is taken over the space outside the bag labeled $\Xi$ shown in Fig.~\ref{fig:bag}. Therefore, the total baryon number of the bag is $Q = Q_\mathrm{in} + Q_\mathrm{out}$.

As is shown in Eq.~(\ref{FluxV}), the vector meson field gets flux by interacting with $A$ and $A$ describes the monopole density inside the bag in a topological manner. We can restrict the complex scalar field $\Phi$ on the bag surface: $\phi=\Phi|_\Omega$. The field $\phi$ inherits a Higgs-type potential $V(\phi^*\phi)$ from $\Phi$ and minimally couples with $V$. For the one-flavor case, we can write down an effective theory at leading order on the bag surface
\begin{equation}
	\int_\Omega|\mathrm{d}\phi-\mathrm{i}V\phi|^2+\frac{N_c}{4\pi}V\mathrm{d}V-V(\phi^*\phi).
\end{equation}
This is precisely the Chern-Simons-Higgs theory, which has been conjectured to exist on the $\eta^\prime$ domain wall~\cite{Lin:2023qya}, with vortex solutions proposed to describe baryons or multibaryon configurations in 2+1 dimensions. For the multiflavor case, the theory generalizes to a non-Abelian form
\begin{equation}
	\int_\Omega|\mathrm{d}\bm{\phi}-\mathrm{i}V\bm{\phi}|^2+\frac{N_c}{4\pi}{\rm tr}\left(V\mathrm{d}V-i\frac{2}{3}V^3\right)-V(\bm{\phi}^\dagger\bm{\phi}),
\end{equation}
where $\bm{\phi}=(\phi^1,\phi^2,\ldots,\phi^{N_f})^T$ is a complex field with $N_f$ components. Thus, there exists a duality in the theory on the $\eta^\prime$ domain wall, as conjectured in~\cite{Gaiotto:2017yup,Gaiotto:2017tne}
\begin{equation}
	SU(N_c)_{-N_f}+N_f \text{ fermions} \longleftrightarrow U(N_f)_{N_c}+N_f \text{ scalars}.
\end{equation}
Under level-rank duality, the dual $SU(N_c)_{-N_f}$ Chern-Simons theory emerges on the bag surface, which should be identified as gluon field, so the level-rank duality becomes a duality between vector meson and gluon.

Therefore, we conclude that there are actually two Chern-Simons theories present on the bag surface. One is a dynamical $U(N_f)_{-N_c}$ Chern-Simons theory associated with the gauge field $\mathbb{A}$, which serves as a counterterm to prevent charge leakage~\cite{Rho:1983bh}. The other is a $U(N_f)_{N_c}$ Chern-Simons theory associated with the vector meson field $V$, which ensures that the bag carries the correct baryon number and is used to construct a one-flavor baryon~\cite{Komargodski:2018odf,Ma:2019xtx,Lin:2023qya,Karasik:2020zyo,Karasik:2020pwu,Karasik:2022tmd}.

\section{Conclusion and discussion}
\label{sec:concl}

Now, we can present the complete picture of the chiral bag. Inside the bag, there exists an $\eta'$ vacuum where monopoles are confined and described by the topological field. To maintain 1-form gauge invariance, a dynamical Chern-Simons field theory emerges on the bag surface, where $\eta'$ experiences a sharp shift. On the bag surface, the dynamical Chern-Simons field interacts explicitly with $\eta'$, resulting in the Witten effect~\cite{Witten:1979ey}, which endows the bag surface with baryon number charge. This induced baryon number charge is equal in magnitude but opposite in sign to that of the monopoles confined within the bag, effectively canceling the total charge. The leakage of baryon number is accompanied by a corresponding leak of color charge. Under level-rank duality, the Chern-Simons theory also serves to block this color charge leak. Consequently, the Chern-Simons theory can be identified as the counterterm on the bag surface, initially introduced last century to maintain 0-form gauge invariance~\cite{Nielsen:1991bk}. This suggests a potentially profound connection between conventional 0-form and higher-form gauge invariance.

Since the baryon number flowing out of the bag is offset on the bag surface, the total baryon number of the bag becomes $0$. To introduce a nonzero baryon number, we must consider the vector meson field outside the bag. This vector meson field occupies a position nearly identical to that of the dynamical Chern-Simons field, acquiring the same flux on the bag surface. Consequently, the vector meson field also gains baryon number due to the Witten effect~\cite{Witten:1979ey}. The flux of the vector meson extends from the bag surface to infinity or terminates on another chiral bag, corresponding to an antibaryon number. This aligns with the hidden Witten-Zumino interaction term that has been thoroughly investigated. Monopoles inside the bag are characterized by the flux of the Chern-Simons field, and the vector meson field on the bag surface acquires an identical flux by the effective interaction. Restricting the effective interaction to the bag surface leads to a Chern-Simons–Higgs theory,
which is conjectured to exist on the $\eta'$ domain wall, where its vortex solutions are proposed to correspond to baryon or multibaryon structures in $2+1$ dimensions~\cite{Lin:2023qya}. Our investigation of the chiral bag model clarifies how the physics of $\eta'$ flows from $3+1$ dimensions to $2+1$ dimensions.

We observe that, on the bag surface, there exist two vector fields, $\mathbb{A}$ and $V$, both carrying flux but with opposite signs. This results in baryon numbers of equal magnitude but opposite sign canceling each other out. It suggests the possible existence of a meson cloud assembled on the bag surface, consisting of a quark and an antiquark, resembling a dipole structure excited by the monopoles inside the bag. Thus, a baryon can be understood as a monopole surrounded by a meson cloud. If the singularity of the monopoles can be eliminated through certain methods, the meson cloud could be identified as a baryon. In fact, the soliton structure formed from the pion field, famously known as a skyrmion, can be viewed as a baryon. Our description of the chiral bag model provides an explanation for why skyrmions can be interpreted as baryons. Of course, if we wish to include the pion field, the chiral bag boundary condition must change~\cite{Chodos:1974je,Rho:1983bh,Goldstone:1983tu,Dreiner:1988jg}, which will inevitably introduce challenges that we leave for future research.

\section*{Acknowledgments} 

The work of Y.L.M. is supported in part by the National Science Foundation of China (NSFC) under Grant No. 12347103 and Gusu Talent Innovation Program under Grant No. ZXL2024363.

\section*{DATA AVAILABILITY} 

No data were created or analyzed in this study.



\end{document}